# Community Detection in Complex Networks using Genetic Algorithm

Mursel Tasgin and Haluk Bingol
*Department of Computer Engineering*
*Bogazici University, Istanbul, Turkey*

Community structure identification has been an important research topic in complex networks and there has been many algorithms proposed so far to detect community structures in complex networks, where most of the algorithms are not suitable for very large networks because of their time-complexity. Genetic algorithm for detecting communities in complex networks, which is based on optimizing network modularity using genetic algorithm, is presented here. It is scalable to very large networks and does not need any priori knowledge about number of communities or any threshold value. It has $O(e)$ time-complexity where $e$ is the number of edges in the network. Its accuracy is tested with the known Zachary Karate Club and College Football datasets. Enron e-mail dataset is used for scalability test.

## I. INTRODUCTION

Community structure identification has created a great interest among physics and computer society who are focusing on the properties of complex networks like the Internet, social networks, citation networks, food networks, e-mail networks and biochemical networks. A complex network is a representation of a complex system from real life in terms of nodes and edges, where a node is an individual member in the system and an edge is a link between nodes according to a relation in the system [1]. As an example, in a social network, a node represents a person and an edge represents social interaction between two people.

Community structure, which is a property of complex networks, can be described as the gathering of vertices into groups such that there is a higher density of edges within groups than between them [2]. From the definition, the nodes in a community should have more intra-community connections rather than inter-community connections. There has been many methods and algorithms proposed so far to reveal the underlying community structure in complex networks. The algorithms require the definition of community that imposes the limit up to which a group should be considered a community [3]. A community detection algorithm's success in finding communities heavily depends on how it defines a community. Many definitions of community exist in the literature [4]. A quantitative definition, network modularity, proposed by Girvan and Newman [5] has been widely used in recent studies as the quality metric for assessment of partitioning a network into communities:

$$Q=\sum_{i}(e_{ii}-a_i^2) \qquad (1)$$

where $i$ is the index of the communities, $e_{ii}$ is the fraction of edges, that connects two nodes inside the community $i$, to the total number of edges in the network and $a_i$ is the fraction of all the edges whose at least one node in the community $i$ to the total number of edges in the network. Most of the recent algorithms use the network modularity as quality metric like Newman's fast algorithm for detecting communities [6], the algorithm for very large networks [2] and the algorithm using Extremal Optimization [3]. As a quality metric, network modularity calculation needs less computation time, when compared to edge betweenness centrality used in Girvan-Newman (GN) algorithm [7], where the overall algorithm has $O(e^3)$ time-complexity, where $e$ is the number of edges. This makes network modularity a practical measure to use it in large networks.

In this paper, we propose a new community detection algorithm based on genetic algorithm, which tries to find the best community structure by maximizing the network modularity. The algorithm has $O(e)$ time-complexity and does not need any priori knowledge about the number of communities or any threshold value, which makes the algorithm useful in very large real-life network. The algorithm outputs the final community structure as the result and does not impose further processing on the output to find the community structure.

## II. COMMUNITY DETECTION METHODS

There have been many different approaches and algorithms to analyze the community structure in complex networks. The algorithms use methods and principles of physics, artificial intelligence, graph theory and even electrical circuits. One of the most known algorithms proposed so far, Girvan-Newman (GN) algorithm, is based on betweenness centrality, which is first proposed by Freeman [8]. The algorithm is a divisive method and has $O(e^3)$ time-complexity. The algorithm produces a hierarchical structure of network, which is called dendogram. Communities are obtained by cutting the dendograms at some point. Radicchi proposed a similar methodology with GN [9], however used a new metric, edge-clustering coefficient whose computation time is less than GN's betweenness centrality, which decreases Radicchi's time-complexity to $O(e^2)$.

Hierarchical agglomerative method based on network modularity, $Q$, is a fast algorithm for detecting communities [5]. If a community has no within-community edges then the network modularity will be $Q=0$, while putting all the nodes into a single community will give $Q=1$. A good community structure will exhibit a network modularity value between 0.3 and 0.7. The algorithm has $O(n^2)$ time-complexity, which is better than GN algorithm and produces very accurate results as well. After the proposal of this algorithm, Newman, Clauset and Moore [2] proposed an improved version of the algorithm that is suitable for large networks. The new algorithm also produces a dendogram; however provides a method to cut the dendogram at some point: when the highest $\Delta Q_{ij}$ value starts to have negative values, it is time to stop merge operations since further joins will not improve community structure.

Community detection using extremal optimization [3] also uses the network modularity. The algorithm tries to optimize the network modularity, $Q$, using artificial intelligence method in a recursive divisive manner. It starts with one community, representing the whole network and continues until the point from which the modularity $Q$ cannot be improved further.

**Drawbacks of current algorithms.** The current algorithms are successful approaches in community detection. However most of these algorithms have time-complexities that make them unsuitable for very large networks. In addition, some algorithms have data structures like matrices etc., which are hard to implement and use in very large networks. Most of the algorithms also need some priori knowledge about the community structure like number of communities etc., where it is impossible to know these values in real-life networks. Some algorithms need threshold values as well, which is another problem for an algorithm, because of variant nature of different complex networks.

## III. THE ALGORITHM

Our algorithm is based on optimization of network modularity for a predefined number of iterations using genetic algorithm's methods [12]. The network structure is represented in a suitable data structure for genetic algorithm, and nodes are placed in random communities at the beginning of the algorithm. There may be at most $n$ communities, where $n$ is number of nodes in the network.

### Genetic Algorithms

Genetic algorithm first proposed in [10] is an optimization method in artificial intelligence. It is a practical method especially when the solution space of a problem is very large and an exhaustive search for the exact solution is impractical. In genetic algorithm, potential solution members in the solution set should be represented in a suitable data representation. Each member in solution set, which is called a chromosome, represents a possible solution to the problem and the algorithm tries to find the best fitting solution member. In order to improve the quality of the solution members the algorithm uses genetic operations on possible solution members for a predefined number of iterations. The algorithm randomly initializes the chromosomes at the beginning. Then for a number of iterations, it uses a fitness function to assign a fitness value to each solution member, which shows how good a solution member is to solve the problem. It reproduces solution members for new population who will be used in the next iteration, by performing cross-over between members selected according to their fitness values. It also applies some random mutation to members.

Genetic algorithm is a fast algorithm for converging a problem to a smaller solution space, and if it has a good fitness evaluation function, it produces near optimal solutions. The power of the algorithm is its cross-over mechanism, which produces better solution members for next generations.

### The Algorithm

We use the network modularity value as the fitness value for each solution member and modify the steps of genetic algorithm to satisfy the needs of our algorithm; like change in cross-over operation and insertion of some additional steps during initial population creation. The algorithm starts with initial population creation. An integer array (or vector) $a$ is used for data representation of community detection problem. The array stores the community identifier (commID) of nodes, that is $a_i$ is the community identifier of the node $i$. The array has $n$ elements and is called as a *chromosome* in genetic algorithm terms. There are a number of chromosomes holding different community configuration information in the population. See Figure 1 for the data representation of a chromosome for the algorithm.

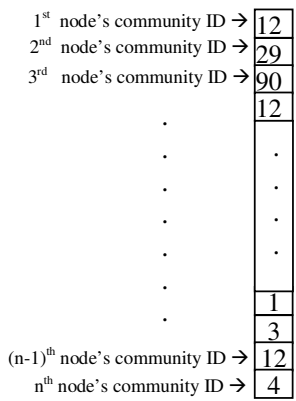

**Figure 1.** Chromosome representation of the algorithm

During initial population creation, each node is assigned a random community identifier. However we need a mechanism to give a bias for initial placement of nodes into communities. If two nodes are to be in the same community, they should have connectivity with each other; in the simplest case they might be neighbors. From this assumption, after assigning random community IDs to nodes, we randomly select some nodes and assign their community IDs to all of their neighbors. This bias in the initial population creation improves the convergence of the algorithm and eliminates unnecessary iterations.

After initial population creation, we perform genetic operations for a number of iterations. During each iteration, the algorithm evaluates fitness values of each solution member, performs cross-over between members, performs mutation and the population becomes ready for next iteration. In each iteration, we sort the solution members according to their fitness value and keep a number of solution members at the top of the list for next generations, which is the *preserving of better genomes* in genetic algorithm. By this way (survival of the fittest), it is guaranteed that a good chromosome, whenever it is created, is never lost.

**Cross-over.** Cross-over operation in genetic algorithm is done by selecting two chromosomes according to their fitness values. Then a cross-over point in chromosome is selected, all the genes after that selection point is exchanged between chromosomes. See Figure 2 for cross-over.

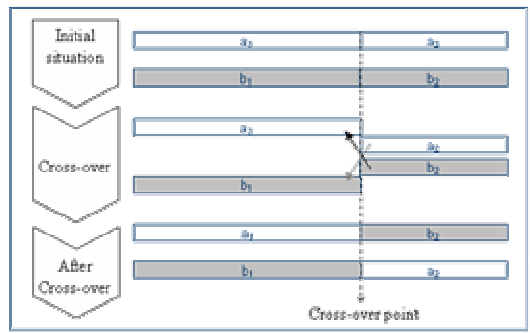

**Figure 2.** Cross-over in genetic algorithm

We modify the cross-over operation in our algorithm. We do not simply exchange genes; instead transfer the community identifiers of nodes in a community to nodes in the destination chromosome. As community structure is a relational property and different community identifiers in different chromosomes may mean the same community. For example, communityID=1 in solution member A and communityID=34 in solution member B have identical members, while communityID=1 in solution B has nothing to do with communityID=1 in solution A

The detail of the cross-over in our algorithm is as follows: we name the chromosomes taking place in cross-over as *source* and *destination*. Then we randomly select a community from *source* chromosome. We iteratively search the nodes that belong to this community and transfer the community identifiers of those nodes in *source* chromosome to nodes in *destination* chromosome. Modified cross-over operation guarantees the exchange of communities at least in one direction. See Figure 3 for the details of one-way cross-over scheme in our algorithm.

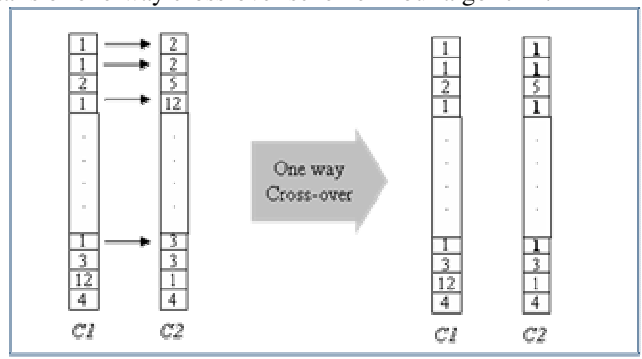

**Figure 3.** Cross-over in community detection algorithm

**Mutations.** After performing a number of cross-overs, we perform mutation to some number of randomly selected chromosomes. In mutation function, a node is placed into a random community in the network.

**Clean-up.** A mechanism to improve the quality of the community splits is needed to reduce the number of nodes that are placed in wrong communities. If the number of such misplacements is high, it is detected by the mechanisms of genetic algorithm via fitness evaluation. However, although the overall fitness value is good for a community split, there may be a small number of misplaced nodes that does not affect the overall fitness value very much. The clean-up process, which is based on a new metric named community variance, aims to reduce all such misplacements.

**Community Variance.** A community should contain more internal links among nodes inside the community than external links with other communities. For this reason, the neighbors of a node should mostly be inside the same community with it. We defined community variance *CV(i)* of a node *i* as the number of different communities among the neighbors and the node itself. *CV(i)* should be low for a good community structure.

$$CV(i) = \frac{\sum_{(i,j) \in E} f_{(i,j)}}{deg(i)}$$
where
$$f_{(i,j)} = \begin{cases} 1, & comm(i) \neq comm(j) \\ 0, & otherwise \end{cases} \quad (2)$$
$deg(i)$ is the degree of vertex $i$
$E$ is the set of edges
$comm(i)$ is the community of vertex $i$

The clean-up process analyzes the community ID variance of some randomly selected nodes. If the community ID variance of that node is larger than threshold value, which is a constant threshold calculated after long batches, then the node and all of its neighbors are put into same community. The new community will be the most common community among the neighbors, the community that contains the highest number of nodes in the neighborhood of the selected node. If threshold value is not exceeded, no operation is performed on community IDs. This process both improves the quality of the community division by eliminating wrongly placed nodes due to random behaviors of the algorithm and also provides a better mechanism to identify the communities of hub nodes, who reside between communities.

At the end of the iterations, the algorithm finishes the optimization process of finding the best community partition. The nodes in the network are assigned a unique community identifier during initial population creation and the commID's of nodes change during the optimization processes of the algorithm. In the end, the algorithm lists each node identifier and their corresponding community identifiers. The final output is the resultant community structure and the algorithm doesn't impose a post-process. The algorithm has *O(e)* time-complexity, where *e* is the number of edges. The fitness evaluation function is the most time-consuming process in the algorithm. The iteration count and number of the chromosomes in the population are directly affecting the performance of the algorithm. However increasing the population size or iteration count does not yield better results after some point. These values do not increase the time-complexity of the algorithm.

## IV. EXPERIMENTAL RESULTS

In our experiments we used a value between 200 and 500 as iteration count and a value between 100 and 250 as population size. We tuned the parameters of genetic algorithm by analyzing the algorithm for long batches and did not change those values after that tuning. We tested the accuracy of the algorithm on two well-known data sets, namely the Zachary Karate Club and the College Football datasets and its scalability on a new data set, Enron e-mail network dataset.

### Zachary Karate Club

The Zachary Karate Club data contains the community structure in a karate club, which is analyzed first in [11]. The network consists of 34 vertices and 78 edges. We run our algorithm on this dataset for a number of times. The algorithm finds 97%-100% correct community structure. It sometimes places the node 10 into wrong community.

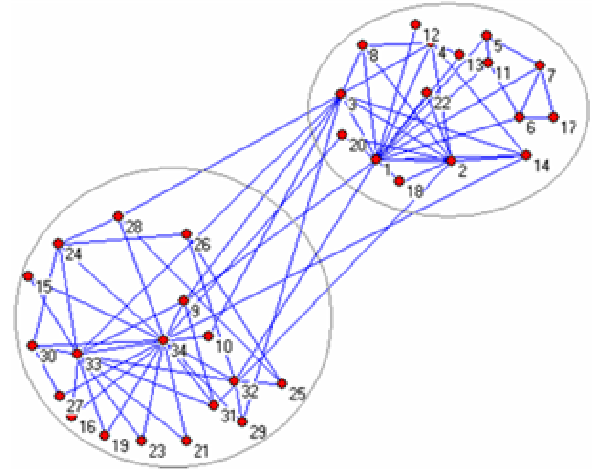

**Figure 4.** Real community structure of Zachary Karate Club

### College Football Network

College football network is composed by the college football matches in USA, for Division I during 2000. The nodes in the network are the college football teams and there is a link between two teams if they played a match. The real community structure is the conferences that each team belongs to. The teams tend to play more matches with teams that are in the same conference and play less inter-conference matches. On the average 6 intra-conference matches and 3 inter-conference matches played by each single team.

The actual data is gathered from web, however all the teams did not satisfy the rule that number of intra-conference matches should be more than inter-conference matches. For this reason, the teams that played more inter-conference matches are removed from the network because they do not exhibit a good community structure in data. The resultant dataset contains 93 teams and 452 matches among these teams. There are 10 conferences and these conferences show the actual community structure of the network.

The algorithm is run on this network and produces 93% accurate community structure in this network. It finds the 100% correct community structure during some runs as well. The fast community detection algorithm [6] is also run on this network, however it produces 78% accurate community structure and it needs the number of communities to produce community structure

**Enron E-mail Network**

Enron, the popular energy company, collapsed as a result of some misleading investment and auditing frauds. After the investigations, mailboxes of 150 important employees were made public for academic purposes. Enron e-mail dataset was available at the end of 2004. Although some recent research has been done on this dataset, in terms of complex networks [13], as far as we know there has been no community detection study on this network yet. Considering the company's organization and some illegal actions are known, the real community structure should be very important for the investigation of the case.

We preprocessed about 512,000 text files to form a complex network dataset, where nodes are the e-mail addresses of the people and the edges are the e-mail connectivity between people. The network consists of 93,526 vertices and 344,264 edges. We tested our algorithm in this network for scalability purposes. We run our algorithm and "the algorithm for very large networks" of [2] on Enron e-mail network since this algorithm is also scalable to this large network as our algorithm. Our algorithm was about 40 to 50 times faster than the first algorithm. Our algorithm run in 25 minutes compared to 23 hours of the other algorithm.

## V. CONCLUSIONS

In this paper, we have proposed a new community detection algorithm, which tries to optimize network modularity using genetic algorithm methods. The contributions of this work are i) applications of genetic algorithm to community detection problem, ii) a community detection algorithm suitable for very large networks, iii) which does not require apriori knowledge of the number of communities, iv) which does not require any threshold value, v) a large data set for community detection studies, and vi) a new metric called community variance.

The algorithm is fast and scalable to very large networks due to its O($e$) time-complexity where $e$ is the number of edges in the network. Unlike most of the previous algorithms, the new algorithm does not need any a priori knowledge about the community structure, like number of communities in the network or any threshold value. It directly produces the community structure of the network in its results without any dendograms. No further processing is needed. We have tested our algorithm's accuracy on well-known network datasets. We also examined the scalability of the algorithm by testing it on Enron e-mail network, which has about 93,000 nodes.

This work was partially supported by Bogazici University Research Projects under the grant number 05A105.